\documentclass[preprintnumbers,article,amsmath,amssymb,floatfix,10pt,prd,onecolumn,
superscriptaddress,nofootinbib]{revtex4-2}
\usepackage{bm}
\usepackage{amsfonts}
\usepackage{latexsym}
\usepackage[latin1]{inputenc}
\usepackage{graphicx}
\usepackage{amsmath}
\usepackage{palatino}
\usepackage{mathpazo}
\usepackage{textcomp}
\usepackage{float}
\usepackage{booktabs}
\usepackage{dcolumn}
\usepackage{ragged2e}
\usepackage{hyperref}
\hypersetup{colorlinks,citecolor=blue}
\hypersetup{colorlinks=true,linkcolor=red,filecolor=magenta,    urlcolor=blue}
\usepackage{amsmath}
\usepackage{xcolor}
\usepackage{orcidlink}
\usepackage{epsfig}
\usepackage[caption=false]{subfig}
\usepackage{commath}
\def\jnl@style{\it}
\def\aaref@jnl#1{{\jnl@style#1}}

\def\aaref@jnl#1{{\jnl@style#1}}

\def\aj{\aaref@jnl{AJ}}                   
\def\apj{\aaref@jnl{ApJ}}                 
\def\apjl{\aaref@jnl{ApJ}}                
\def\apjs{\aaref@jnl{ApJS}}               
\def\apss{\aaref@jnl{Ap\&SS}}             
\def\aap{\aaref@jnl{A\&A}}                
\def\aapr{\aaref@jnl{A\&A~Rev.}}          
\def\aaps{\aaref@jnl{A\&AS}}              
\def\mnras{\aaref@jnl{Mon.~Not.~Roy.~Astron.~Soc.}}             
\def\prd{\aaref@jnl{Phys.~Rev.~D}}        
\def\prc{\aaref@jnl{Phys.~Rev.~C}}  
\def\prl{\aaref@jnl{Phys.~Rev.~Lett.}}    
\def\qjras{\aaref@jnl{QJRAS}}             
\def\skytel{\aaref@jnl{S\&T}}             
\def\ssr{\aaref@jnl{Space~Sci.~Rev.}}     
\def\zap{\aaref@jnl{ZAp}}                 
\def\nat{\aaref@jnl{Nature}}              
\def\aplett{\aaref@jnl{Astrophys.~Lett.}} 
\def\apspr{\aaref@jnl{Astrophys.~Space~Phys.~Res.}} 
\def\physrep{\aaref@jnl{Phys.~Rep.}}      
\def\physscr{\aaref@jnl{Phys.~Scr}}       
\def\commat{\aaref@jnl{Comm.~Math.~Phys.}}              
\def\science{\aaref@jnl{Science}}               
\def\cqg{\aaref@jnl{Classical Quant.~Grav.}}            
\def\jpcs{\aaref@jnl{JPCS}}                                     
\def\ijmpd{\aaref@jnl{Int.~J.~Mod.~Phys.~D}}                    
\def\grg{\aaref@jnl{Gen.~Relat.~Gravit.}}               
\def\rpp{\aaref@jnl{Rep.~Prog.~Phys.}}          
\def\npa{\aaref@jnl{Nucl.~Phys.~A}}        
\def\lrr{\aaref@jnl{Living Rev.~Rel.}}                   
\def\jcap{\aaref@jnl{J.~Cosmology Astropart.~Phys.}}    
\def\rmp{\aaref@jnl{Rev.~Mod.~Phys.}}   
\def\epjc{\aaref@jnl{Eur.~Phys.~J.~C}}
\def\plb{\aaref@jnl{~Phy.~Lett.~B}}
\def\mpla{\aaref@jnl{Mod.~Phy.~Lett.~A}}
\def\arxiv{\aaref@jnl{arxiv.org}}

\allowdisplaybreaks[1]

\begin{document}
\color{black}       
\title{Quantum Gravity Evolution in the Hawking Radiation of a Rotating Regular Hayward Black Hole}
\author{Riasat Ali}
\email{riasatyasin@gmail.com}
\affiliation{Department of Mathematics, GC,
University Faisalabad Layyah Campus, Layyah-31200, Pakistan}

\author{Rimsha Babar}
\email{rimsha.babar10@gmail.com}
\affiliation{Division of Science and Technology, University of Education, Township, Lahore-54590, Pakistan}

\author{P.K. Sahoo\orcidlink{0000-0003-2130-8832}}
\email{pksahoo@hyderabad.bits-pilani.ac.in}
\affiliation{Department of Mathematics, Birla Institute of Technology and
Science-Pilani,\\ Hyderabad Campus, Hyderabad-500078, India.}
%
\begin{abstract}
In this paper, we study two different phenomena
(the Newman-Janis algorithm and the semiclassical Hamilton-Jacobi method) to analyze the Hawking temperature ($T_H$) for massive $4$-dimensional regular Hayward BH with spin parameter. First of all, we compute the rotating regular Hayward black hole solution by taking the Newman-Janis algorithmic rule. We derive the $T_H$ for rotating regular Hayward BH with the help of surface gravity.
We have also analyzed the effects of spin parameter $a$ and free parameter $l$ on $T_H$ with the help of graphs.
Moreover, we investigate the quantum corrected Hawking temperature ($T'_H$) for rotating regular Hayward black hole. To do so, we utilize the Lagrangian filed equation in the background of GUP within the concept of WKB approximation and semiclassical
Hamilton-Jacobi method. The $T'_H$ of rotating regular Hayward BH depends upon correction parameter $\beta$, BH mass $m$, spin parameter $a$, free parameter $l$ and BH radius $r_+$. We also study the graphical behavior of $T'_H$ versus event horizon $r_+$ for rotating regular Hayward BH and check the influences of quantum gravity parameter $\beta$, spin parameter $a$ and free parameter $l$ on the stability of corresponding black hole. Moreover, we study the significance's of logarithmic entropy correction for regular rotating Hayward BH.

\end{abstract}

\date{\today}

\maketitle

\section{Introduction}
The first idea of Black Hole (BH) was propounded by John Michell ($1783$).
Black Holes are the most imperative discovery of the universe. Moreover, BH physics preserve the
mystery of the present in the form of information paradox \cite{x1} but disregarding all
profound investigations on BH physics the singularity at center is also an open problem until we
introduce quantum gravity theory \cite{x2}. In order to solve the singularity issues many various solutions
of BHs were introduced and these non-singular solutions are known as regular BHs.
In recent years these models pulled a lot of attraction of researchers, particularly the
solution of non-linear electrodynamics with theory of Einstein gravity. By using the non-linear
electrodynamics many Bardeen like regular BHs were obtained to remove the singularities \cite{x3}.
Furthermore, the evaporation and quantum corrections have also been discussed for regular BHs \cite{x4, x5, x45}.
Hayward \cite{x6} obtained the same type of regular BH solution with well-defined asymptotic
limits such as for $r\rightarrow\infty$ it becomes Schwarzschild and de-Sitter when $r\rightarrow0$.
Halilsoy and his colleagues \cite{x7} have revisited the regular Hayward BH and constructed
a thin shell wormhole by using the different equations of state. Molina and  Villanueva \cite{x8} have
derived the roots and studied the thermodynamics of Hayward BH.
In order to study the thermodynamics of BHs many approaches have been introduced.
Many authors have investigated the $T_H$ for various BHs by using the
semi-classical tunneling method \cite{x9}-\cite{x23}. As a comparison between Hawking computation
and tunneling strategy, it is not difficult to see that the Hawking technique is an immediate strategy
however its complexity to speculation to any remaining space times is failed while the tunneling
methods have been effectively applied to a wide scope of both the BH event and cosmological horizons.
Lately, the generalized uncertainty principle (GUP) has been the subject of numerous curiously
works and many research have been appeared in which the standard uncertainty principle is
generalized as microphysics framework. To investigate the quantum gravity effects on the
tunneling strategy, it is fascinating to associate the tunneling analysis with a observable minimal length in the following expression
\begin{equation}
\triangle\mathfrak{p}\triangle x\geq\frac{\hbar}{2}\Big[1+(\triangle\mathfrak{p})^2\beta\Big],
\end{equation}
here $\beta=\frac{\beta_0}{M^2_p}$, $\beta_0$ and $M_p$ denotes the dimensionless parameter and Plank mass, respectively.
The computation of the above equation depends upon the generalized standard commutation relation $[x_u, \mathfrak{p_v}]=i\hbar\delta_{uv}[1+\beta\mathfrak{p}^2]$, here $x_u$ and $\mathfrak{p}_v$ stands for position
and momentum operators, respectively.
Ali and his colleagues \cite{x24} have investigated the gravity effects on Hawking radiations
for charged black strings via Rastall gravity.

The modifications of the usual exchange relations is not unique.
Many modifications of exchange relations are suggested to \cite{x25}-\cite{x27}.
To get some information about the property of the gravity of quantum,
these corrections are considered widely. Black holes are valuable objects for researching the impacts of gravity. Some fascinating outcomes and discoveries were acquired by
the formation of the quantum of gravity impacts into BH theory via GUP \cite{x23}.
The GUP effects have been analyzed to massive vector and scalar particles from warped DGP gravity BH \cite{x28}.
Ling \cite{x29} has studied the tunneling approach for fermions particles from black lenses in $5$D.
As a result of their investigation they concluded that by using the tunneling approach one can calculate the correct values of $T_H$ for rotating BHs.

The quantum corrections for charged particles via tunneling method for modified Reissner
BH have been investigated \cite{x30}. Gecim and Sucu \cite{x31} have studied the GUP effects
on Hawking radiation in the background of $(2 + 1)$-dimensional Warped-Ad$S_3$.

Cimdiker and his colleagues \cite{v1} have discussed the $4$-dimensional BH in symmergent gravity and showed that the horizon radius, Hawking temperature,
Bekenstein-Hawking entropy, photon deflection angle and shadow angular radius are sensitive
investigations of the symmergent gravity of the fundamental quantum field theory.
 \"{O}vg\"{u}n \cite{v2} has studied the solution of an exact confining charged BH to the scalar-tensor representation
of regularized $4$-dimensional Einstein-Gauss-Bonnet gravity and investigated the BH thermodynamics and physical properties 
for the corresponding BH i.e., Hawking temperature, specific heat, quasinormal modes and BH shadow.
Okay and  \"{O}vg\"{u}n \cite{v3} have investigated the effects of nonlinear electrodynamics on non-rotating BHs,
parametrized by magnetic charge and the field coupling parameter. They also discussed the 
Hawking temperature and heat capacity for this BH solution. 
Sakalli and \"{O}vg\"{u}n \cite{v4}-\cite{v6} have studied the Hawking radiations phenomenon 
by using quantum tunneling method for spin-$1$ particles from Non-stationary Metrics, Rindler 
modified Schwarzschild BH as well as Lorentzian Wormholes in $3+1$ dimensions.

The BH entropy was suggested by Hawking paper that the BH horizon area never decreases \cite{en1}, as well as the evolution of this solution into
the four laws of BH mechanics \cite{en2}.

The main purpose of this paper is to check the quantum gravity effects for regular rotating Hayward BH
and to discuss the stability condition of BH in the presence/absence of the quantum of gravity effects with
the help of graphical analysis. This article is established in the following manner: In Sec. \textbf{2},
we analyze a metric for regular rotating Hayward BH by considering the Newman-Jannis algorithmic rule and also compute
the $T_H$ for derived BH metric. Section \textbf{3} depicts the graphical behavior of
$T_H$ versus event horizon $r_+$ and describes the stability of corresponding BH.
Section \textbf{4} investigates the $T'_H$ for regular rotating Hyward BH.
Section \textbf{5} premises the influences of quantum gravity parameter $\beta$, spin parameter $a$
and free parameter $l$ on regular rotating Hayward BH.
with graphs.
At last, Sec. \textbf{6} discusses the main results and conclusions.

\section{Regular Hayward Black Hole with Rotation Parameter}
The line element of spherically symmetric static Hayward non-singular BH
enclosed in \cite{x7} is given as
\begin{equation}
ds^{2}=-F(r)dt^2+\frac{1}{F(r)}dr^2+r^2 d\theta^2+r^2 \sin^2\theta d\phi^2,\label{d1}
\end{equation}
where $F(r)=\Big(1-\frac{2mr^2}{r^3+2ml^2}\Big)$,
$m$ and $l$ represents the two free parameters.

By applying the Newman-Janis algorithm, we derive the regular Hayward
BH metric with rotation parameter for rotating case. So, through the transformation of
$(t, r, \theta, \phi)$ to $(u, r, \theta, \phi)$ coordinates, we obtain
\begin{eqnarray}
du=dt-\frac{dr}{F}.\label{A}
\end{eqnarray}
After using the above transformation the metric (\ref{d1}) can be expressed as follows
\begin{equation}
ds^{2}=-F(r)du^2-2dudr+r^2 d\theta^2+r^2 \sin^2 \theta d\phi^2.
\end{equation}
In the null tetrad framework the metric can be written as
\begin{eqnarray}
g^{\mu\nu}=-l^\nu n^\mu-l^\mu n^\nu+m^\mu \bar{m}^{\nu}+m^\nu \bar{m}^{\mu}.
\end{eqnarray}
The corresponding components are given as
\begin{eqnarray}
l^{\mu}&=&\delta_{r}^{\mu},~~~n^{\nu}=\delta_{u}^{\mu}-\frac{1}{2} F \delta_{r}^{\mu},\nonumber\\
m^{\mu}&=&\frac{1}{\sqrt{2}r} \delta_{\theta}^{\mu}+\frac{i}{\sqrt{2}r \sin\theta}\delta_{\phi}^{\mu},\nonumber\\
\bar{m}^{\mu}&=&\frac{1}{\sqrt{2}r} \delta_{\theta}^{\mu}-\frac{i}{\sqrt{2}r \sin\theta}\delta_{\phi}^{\mu}.\nonumber
\end{eqnarray}
For any point in the BH space-time the null vectors of the null tetrad satisfy the relations
$l_{\mu}l^{\mu}=n_{\mu}n^{\mu}=m_{\mu}m^{\mu}=l_{\mu}m^{\mu}=m_{\mu}m^{\mu}=0$
and $l_{\mu}n^{\nu}=-m_{\mu}\bar{m}^{\mu}=1$ in the $(u, r)$ plane the coordinates transformation are
$u\rightarrow u(real~part)-ia\cos\theta$(imaginary part), $r\rightarrow r(real~part)+ia\cos\theta$(imaginary part), then we perform the transformation
$F(r)\rightarrow\tilde{F}(a, r, \theta)$ and $\Sigma^2=a^2 \cos^2\theta+r^2$.
The vectors in the $(u, r)$ space become
\begin{eqnarray}
l^{\mu}&=&\delta_{r}^{\mu},~~~n^{\nu}=\delta_{u}^{\mu}-\frac{1}{2}
\tilde{F} \delta_{r}^{\mu},\nonumber\\
m^{\mu}&=&\frac{1}{\sqrt{2}r}\left(\delta_{\theta}^{\mu}+ia \sin\theta\big(\delta_{u}^{\mu}
-\delta_{r}^{\mu}\big)+\frac{i}{\sin\theta}\delta_{\phi}^{\mu}\right),\nonumber\\
\bar{m}^{\mu}&=&\frac{1}{\sqrt{2}r}\left(\delta_{\theta}^{\mu}-ia\sin\theta\big(\delta_{u}^{\mu}
-\delta_{r}^{\mu}\big)-\frac{i}{\sin\theta}\delta_{\phi}^{\mu}\right),\nonumber
\end{eqnarray}
From the definition of null tetrad the metric tensor $g^{\mu r}$ in
the $(u, r, \theta, \phi)$ coordinates are given by
\begin{eqnarray}
g^{uu}&=&\frac{a^2 \sin^2\theta}{\sum^2},~~~g^{ur}=g^{ru}=-1-\frac{a^2 \sin^2\theta}{\sum^2}
,~~~g^{rr}=\tilde{F}+\frac{a^2 \sin^2\theta}{\sum^2},~~~
g^{\theta\theta}=\frac{1}{\sum^2},\nonumber\\
g^{\phi\phi}&=&\frac{1}{\sum^2 \sin^2\theta},~~~g^{u\phi}=g^{\phi u}=\frac{a}{\sum^2},~~~
g^{r\phi}=g^{\phi r}=-\frac{a}{\sum^2},\nonumber
\end{eqnarray}

The new line element in the tetrad transformation can be specified by
\begin{eqnarray}
ds^{2}=-\tilde{F}(r)du^2-2a\sin ^2\theta\big(1-\tilde{F}\big)du d\phi-2dudr+2a \sin ^2\theta drd\phi+\Sigma^2d\theta^2
+\sin^2\theta\Big(\Sigma^2-a^2\big(\tilde{F}-2\big)\sin^2\theta\Big)d\phi^2.
\end{eqnarray}
Finally, we perform the transformation from coordinates $(u, r , \theta, \phi)$ to
$(t, r, \theta, \phi)$ as
\begin{equation}
du=dt+\lambda(r)dr,~~~d\phi=d\phi+Z(r)dr,
\end{equation}
where
\begin{equation}
\lambda(r)=\frac{a^2+r^2}{r^2\tilde{F} +a^2},~~~~~~ Z(r)=-\frac{a}{a^2+r^2\tilde{F}}
,~~~\tilde{F}(r, \theta)=\frac{r^2F(r)+a^2 \cos^2\theta}{\Sigma^2}.\nonumber
\end{equation}
The regular Hayward BH metric with new spin (rotation) parameter in $(u, r, \theta, \phi)$ can be derived in the following form
\begin{eqnarray}
ds^{2}&=&-\left(1-\frac{2mr^4/r^3+2ml^2}{\Sigma^2}\right)dt^2
-2a\left(\frac{2mr^4/r^3+2ml^2}{\Sigma^2}\right)\sin^2\theta dt d\phi+\frac{\Sigma^2}{\Delta}dr^2+\Sigma^2d\theta^2\nonumber\\
&+&\sin\theta^2\left[\Sigma^2-a^2\left(\frac{2mr^4/r^3+2ml^2}{\Sigma^2}\right)\sin\theta^2\right]d\phi^2.\label{m1}
\end{eqnarray}
where
\begin{eqnarray}
\Delta&=&r^2+a^2-\frac{2mr^4}{r^3+2ml^2}.\nonumber
\end{eqnarray}
Now, we discuss the $T_{H}$ from regular
Hayward BH for rotating case with the help of surface gravity by using the given formula
\begin{eqnarray}
T_{H}=\frac{\kappa}{2\pi},~~~~~\kappa=\frac{\Delta'_r}{2(r^2_+ + a^2)},
\end{eqnarray}
The $T_{H}$ from regular Hayward BH for rotating case can be derived as
\begin{eqnarray}
T_{H}&=&\frac{2r_+\Big(2m l^2+r_+^3\Big)^2+6mr_+^6-8mr_+^3\Big(2m l^2 +r_+^3\Big)}
{4\pi\Big(2ml^2+r_+^3\Big)^2\Big(r_+^2+a^2\Big)}.\label{bb}
\end{eqnarray}
The above $T_{H}$ depend upon the mass $m$ of BH, free parameter $l$ and rotation parameter $a$.

\section{Graphical Analysis of Hawking Temperature $T_{H}$ for Regular Rotating Hayward BH}
This part investigate the behavior of $T_{H}$
versus event horizon $r_+$ with the help of plots for regular rotating Hayward BH. We study the effects of arbitrary
parameter $l$ as well as spin parameter $a$ for $T_{H}$ of regular rotating Hayward BH.
We analyze the behavior of $T_{H}$ by choosing the fixed
value of mass $m=1$. We also investigate the stability
of regular rotating Hayward BH under the influence of different parameters.

\begin{center}
\includegraphics[width=8cm]{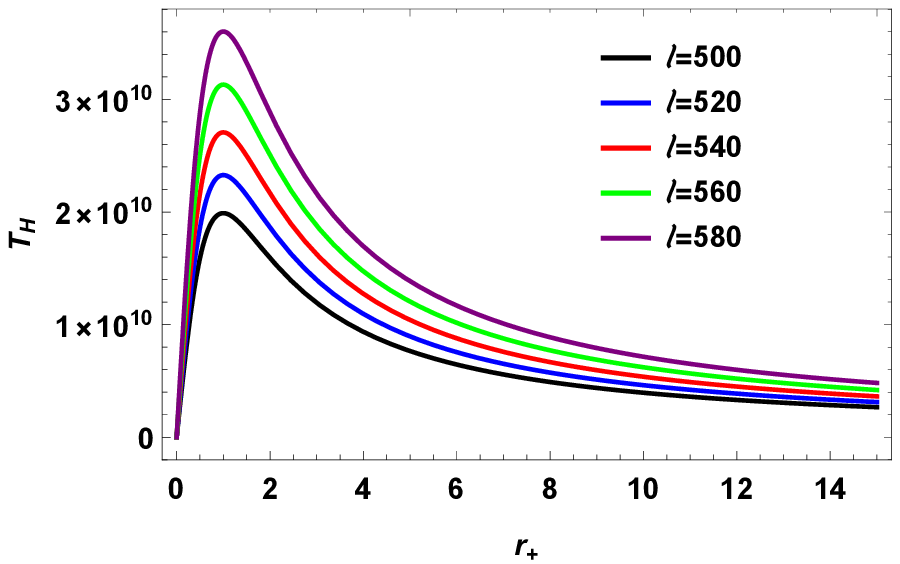}\includegraphics[width=8cm]{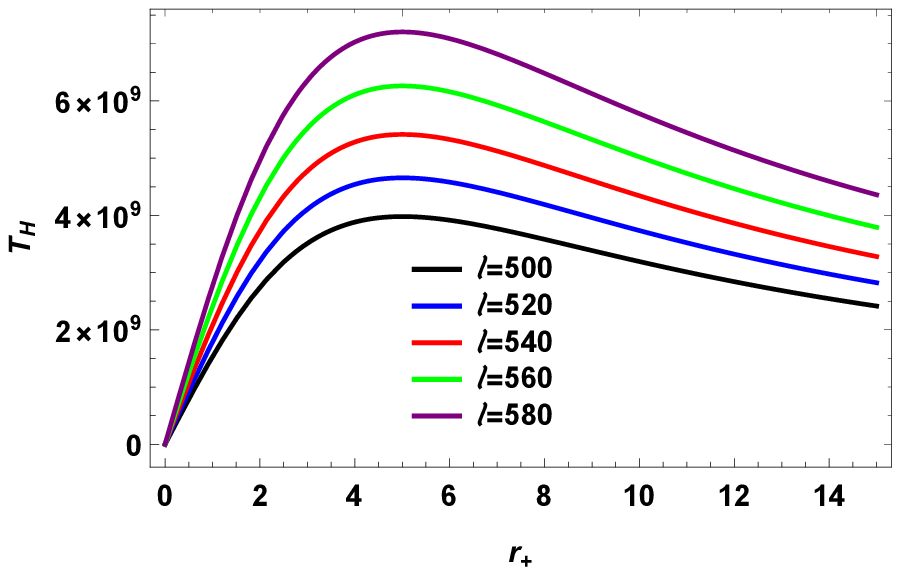}\\
{Figure 1: Hawking temperature $T_{H}$ via horizon $r_{+}$.}
\end{center}

\textbf{Fig. 1}: states the behavior of $T_{H}$ for different values of $l$ and for fixed values of spin parameter
$a=1$ and $a=5$, respectively.
The left hand plot shows the $T_{H}$ after getting a maximum height eventually goes down and obtain an asymptotically
flat case till $r_{+}\rightarrow\infty$. This case ensures the stable form of BH.
We can also test that the $T_{H}$ increases with the increasing value of $l$.

The right hand plot shows the presentation of $T_{H}$ for changing values of $l$ as well as fixed value
of spin parameter $a=5$. One can observe the $T_{H}$ decreases with the increasing values of $r_+$.
This behavior with positive high $T_{H}$ shows the stability of BH.
\begin{center}
\includegraphics[width=8cm]{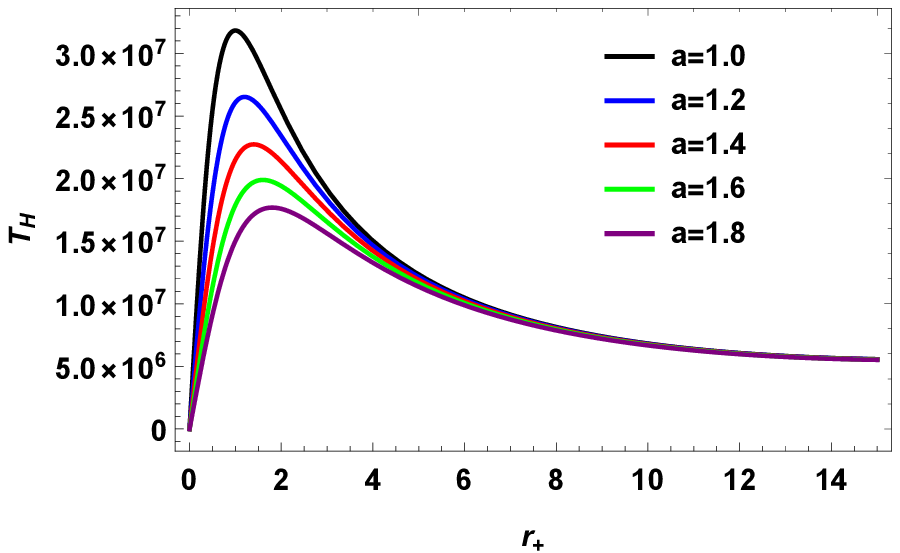}\includegraphics[width=8cm]{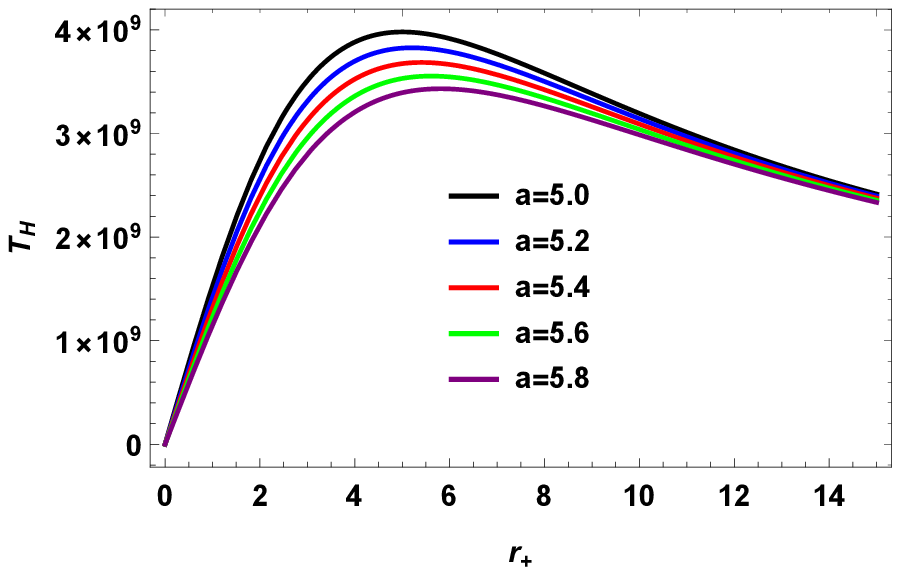}\\
{Figure 2: Hawking temperature $T_{H}$ via horizon $r_{+}$.}
\end{center}
\textbf{Fig. 2}: gives the behavior of $T_{H}$ for varying values of spin parameter $a$
and fixed values of arbitrary constant $l=100$ and $l=500$, respectively.
The left hand plot shows the presentation of $T_{H}$ for fixed value of $l=100$.
By the emission of Hawking radiation, the $T_{H}$ reaches at a maximum height
at the final stage of the evaporation and afterward suddenly drops to zero so that a
stable remnant is showed up.

The right hand plot depicts the behavior of $T_{H}$ for different values of $a$
and fixed value of $l=500$.
The increasing values of $r_{+}$ shows a decreasing behavior of $T_{H}$ after
a maximum height which shows the physical and stable behavior of BH. The $T_{H}$
decreases when we increase the values of $a$.

\section{Quantum corrected temperature for Regular Rotating Hayward BH }
The metric Eq. (\ref{m1}) can be re-written as
\begin{eqnarray}
ds^{2}&=&-Adt^{2}+Bdr^{2}+Cd\theta^{2}
+D dy^{2}+Edt dy,\label{aa}
\end{eqnarray}
here
\begin{eqnarray}
A&=&\left(1-\frac{2mr^4/r^3+2ml^2}{\Sigma^2}\right),~~~B=\frac{\Sigma^2}{\Delta},
~C=\Sigma^2,~~~
D=\sin\theta^2\left[\Sigma^2-a^2\left(\frac{2mr^4/r^3+2ml^2}{\Sigma^2}\right)\sin\theta^2\right],\nonumber\\
~E&=&2a\left(\frac{2mr^4/r^3+2ml^2}{\Sigma^2}\right)\sin^2\theta.\nonumber
\end{eqnarray}
Lagrangian field equation with quantum gravity parameter is given by
\begin{eqnarray}
\partial_{\mu}\Big(\sqrt{-g}\phi^{\nu\mu}\Big)+\sqrt{-g}\frac{m^2}{\hbar^2}\phi^{\nu}
+\hbar^{2}\beta\partial_{0}\partial_{0}\partial_{0}\Big(\sqrt{-g}g^{00}\phi^{0\nu}\Big)
-\hbar^{2}\beta\partial_{i}\partial_{i}\partial_{i}\Big(\sqrt{-g}g^{ii}\phi^{i\nu}\Big)=0,\label{m2}
\end{eqnarray}
where $\phi^{\nu\mu}$ shows the anti-symmetric
tensor, $g$ denotes the coefficient matrix determinant and $m$ represents the
mass of radiated particle. The $\phi^{\nu\mu}$ tensor is calculated by the given formula
\begin{equation}
\phi_{\nu\mu}=\Big(1-{\hbar^2\beta\partial_{\nu}^2}\Big)\partial_{\nu}\phi_{\mu}-
\Big(1-{\hbar^2\beta\partial_{\mu}^2}\Big)\partial_{\mu}\phi_{\nu},
\end{equation}
where $\beta$ is a parameter of quantum gravity.

The non-zero elements of $\phi^{\nu\mu}$ are given as
\begin{eqnarray}
&&\phi^{0}=\frac{-D\phi_{0}+E\phi_{3}}{E^2+AD},~~~\phi^{1}=\frac{1}{B}\phi_{1},
~~~\phi^{2}=\frac{1}{C}\phi_{2},~~~
\phi^{3}=\frac{E\phi_{0}+A\phi_{3}}{E^2+AD},~~~\phi^{01}=\frac{-D\phi_{01}
+E\phi_{13}}{B\Big(E^2+AD\Big)},~~~\phi^{12}=\frac{1}{BC}\phi_{12},\nonumber\\
&&\phi^{02}=\frac{-D\phi_{02}}{C\Big(E^2+AD\Big)},
~~~\phi^{03}=\frac{\Big(A^2-AD\Big)\phi_{03}}{\Big(E^2+AD\Big)^2},~~~
~\phi^{13}=\frac{1}{BAD+E^2}\phi_{13},~~\phi^{23}=\frac{E\phi_{02}+A\phi_{23}}{C\Big(E^2+AD\Big)}.\nonumber
\end{eqnarray}
The WKB approximation is given by
\begin{equation}
\phi_{\nu}=c_{\nu}\exp\left[\frac{i}{\hbar}S_{0}\big(t,r,\theta, y\big)+
\Sigma \hbar^{n}\Theta_{n}\big(t,r, \theta,y\big)\right].
\end{equation}
After putting all the given values in Eq. (\ref{m2}), we get the set of field equation in this form
\begin{eqnarray}
&&\frac{D}{(E^2+AD)B}\Big[c_{1}(\partial_{0}S_{0})(\partial_{1}S_{0})+\beta c_{1}
(\partial_{0}S_{0})^{3}(\partial_{1}S_{0})-c_{0}(\partial_{1}S_{0})^{2}
-\beta c_{0}(\partial_{1}S_{0})^4\Big]-\frac{E}{B(E^2+AD)}\Big[c_{3}(\partial_{1}S_{0})^2\nonumber\\
&+&\beta c_{3}(\partial_{1}S_{0})^4-c_{1}(\partial_{1}S_{0})(\partial_{3}S_{0})
-\beta c_{1}(\partial_{1}S_{0})(\partial_{3}S_{0})^2\Big]+\frac{D}{C(E^2+AD)}
\Big[c_{2}(\partial_{0}S_{0})(\partial_{2}S_{0})+\beta c_{2}(\partial_{0}S_{0})^3(\partial_{2}S_{0})\nonumber\\
&-&c_{0}(\partial_{2}S_{0})^2-\beta c_{0}(\partial_{2}S_{0})^4\Big]
+\frac{AD}{(E^2+AD)^2}\Big[c_{3}(\partial_{0}S_{0})(\partial_{3}S_{0})
+\beta c_{3}(\partial_{0}S_{0})^{3}(\partial_{3}S_{0})-c_{0}(\partial_{3}S_{0})^{2}
-\beta c_{0}(\partial_{3}S_{0})^4\Big]\nonumber\\
&-&m^2\frac{{D c_{0}}-{E c_{3}}}{(E^2+AD)}=0,\label{m4}\\
&-&\frac{D}{B(E^2+AD)}\Big[c_{1}(\partial_{0}S_{0})^2+\beta c_{1}
(\partial_{0}S_{0})^4-c_{0}(\partial_{0}S_{0})(\partial_{1}S_{0})-\beta c_{0}(\partial_{0}S_{0})
(\partial_{1}S_{0})^{3}\Big]+\frac{E}{B(E^2+AD)}\Big[c_{3}
(\partial_{0}S_{0})(\partial_{1}S_{0})\nonumber\\
&+&\beta c_{3}(\partial_{0}S_{0})(\partial_{1}S_{0})^3-c_{1}(\partial_{0}S_{0})(\partial_{3}S_{0})-\beta c_{1}
(\partial_{0}S_{0})(\partial_{3}S_{0})^{3}\Big]+\frac{1}{BC}\Big[c_{2}(\partial_{1}S_{0})(\partial_{2}S_{0})+\beta c_{2}(\partial_{1}S_{0})(\partial_{2}S_{0})^3-c_{1}(\partial_{2}S_{0})^{2}\nonumber\\
&-&\beta c_{1}(\partial_{2}S_{0})^{4}\Big]+\frac{1}{B(E^2+AD)}\Big[c_{3}(\partial_{1}S_{0})(\partial_{3}S_{0})+\beta c_{3}(\partial_{1}S_{0})(\partial_{3}S_{0})^3-c_{1}(\partial_{3}S_{0})^2
-\beta c_{1} (\partial_{3}S_{0})^{4}\Big]-\frac{m^2 c_{1}}{B}=0,\label{m5}\\
&+&\frac{D}{C(AD+E^2)}\Big[c_{2}(\partial_{0}S_{0})^2+\beta c_{2}
(\partial_{0}S_{0})^{4}-c_{0}(\partial_{0}S_{0})(\partial_{2}S_{0})
-\beta c_{0}(\partial_{0}S_{0})(\partial_{2}S_{0})^3\Big]+\frac{1}{BC}\Big[c_{2}(\partial_{1}S_{0})^2+\beta c_{2}
(\partial_{1}S_{0})^{4}\nonumber\\
&-&c_{1}(\partial_{1}S_{0})(\partial_{2}S_{0})
-\beta c_{1}(\partial_{1}S_{0})(\partial_{2}S_{0})^3\Big]
-\frac{E}{C(E^2+AD)}\Big[c_{2}(\partial_{0}S_{0})(\partial_{3}S_{0})+\beta c_{2}
(\partial_{0}S_{0})^{3}(\partial_{3}S_{0})-c_{0}(\partial_{0}S_{0})(\partial_{3}S_{0})\nonumber\\
&-&\beta c_{0}
(\partial_{0}S_{0})^3 (\partial_{3}S_{0})\Big]
+\frac{A}{C(AD+E^2)}\Big[c_{3}(\partial_{2}S_{0})(\partial_{3}S_{0})+\beta c_{3}
(\partial_{2}S_{0})^{3}(\partial_{3}S_{0})-c_{2}(\partial_{3}S_{0})^2
-\beta c_{2}(\partial_{3}S_{0})^4\Big]\nonumber\\
&-&\frac{m^2 c_{2}}{C}=0,\label{m6}
\end{eqnarray}
\begin{eqnarray}
&+&\frac{(AD)-A^2}{(E^2+AD)^2}\Big[c_{3}(\partial_{0}S_{0})^2+\beta c_{3}
(\partial_{0}S_{0})^4-c_{0}(\partial_{0}S_{0})(\partial_{3}S_{0})-\beta c_{0}(\partial_{0}S_{0})(\partial_{3}S_{0})^{3}\Big]-\frac{D}{C(E^2+AD)}\Big[c_{3}(\partial_{1}S_{0})^2\nonumber\\
&+&\beta c_{3}
(\partial_{1}S_{0})^{4}-c_{1}(\partial_{1}S_{0})(\partial_{3}S_{0})-
\beta c_{1}(\partial_{1}S_{0})(\partial_{3}S_{0})^3\Big]-\frac{E}{C(E^2+AD)}
\Big[c_{2}(\partial_{0}S_{0})(\partial_{2}S_{0})+\beta c_{2}
(\partial_{0}S_{0})^3(\partial_{2}S_{0})\nonumber\\
&-&c_{0}(\partial_{2}S_{0})^{2}
+\beta c_{0}(\partial_{2}S_{0})^4\Big]
-\frac{m^2 (Ec_{0}-Ac_{3})}{(E^2+AD)}=0.\label{m7}\\
\end{eqnarray}
Using separation of variables technique, we can choose
\begin{equation}
S_{0}=-(E-j\omega)t+W(r)+Jy+\nu(\theta),\label{m8}
\end{equation}
where $E$ stands for the energy of particle, $J$ shows the particles angular
momentum.

By using Eq. (\ref{m8}) into set of Eqs. (\ref{m4}-\ref{m7}), we obtain a $4\times4$ matrix equation as follows
\begin{equation*}
K(c_{0},c_{1},c_{2},c_{3})^{T}=0.
\end{equation*}
So, the elements of the above matrix equation are defined as
\begin{eqnarray}
K_{00}&=&\frac{{-D}}{B(E^2+AD)}\Big[W_{1}^2+\beta W_{1}^4\Big]-\frac{D}{C(E^2+AD)}\Big[J^2+\beta J^4\Big]
-\frac{AD}{(E^2+AD)^2}\Big[\nu_{1}^2+
\beta \nu_{1}^4\Big]-\frac{m^2 D}{(E^2+AD)},\nonumber\\
K_{01}&=&\frac{{-D}}{B(E^2+AD)}\Big[(E-j\omega)+\beta (E-j\omega)^3\Big]
+\frac{E}{B(E^2+AD)}+\Big[\nu_{1}+
\beta \nu_{1}^3\Big],\nonumber\\
K_{02}&=&\frac{{-D}}{C(E^2+AD)}\Big[(E-j\omega)+\beta (E-j\omega)^3\Big],\nonumber\\
K_{03}&=&\frac{{-E}}{B(E^2+AD)}\Big[W_{1}^2+\beta W_{1}^4\Big]-
\frac{AD}{C(E^2+AD)^2}\Big[(E-j\omega)+\beta (E-j\omega)^3\Big]
+\frac{m^2E}{(E^2+AD)^2},\nonumber\\
K_{10}&=&\frac{{-D}}{B(E^2+AD)}\Big[(E-j\omega)W_{1}+\beta (E-j\omega)W_{1}^3\Big]
-\frac{m^2}{B},~~~~~K_{21}=\frac{1}{BC}\Big[J+\beta J^3\Big]W_{1},\nonumber\\
K_{11}&=&\frac{{-D}}{B(E^2+AD)}\Big[(E-j\omega)^2+\beta(E-j\omega)^4\Big]
+\frac{E}{B(E^2+AD)}\Big[\nu_{1}+\beta \nu_{1}^3\Big]
(E-j\omega)-\frac{1}{BC}\Big[J^2+\beta J^4\Big]\nonumber\\
&-&\frac{1}{B(E^2+AD)}\Big[\nu_{1}+\beta \nu_{1}^3\Big]
-\frac{m^2}{B},~~~~~\nonumber
K_{12}=\frac{1}{BC}\Big[W_{1}+\beta W_{1}^3\Big]J,\nonumber\\
K_{13}&=&\frac{{-E}}{B(E^2+AD)}\Big[W_{1}+\beta W_{1}^3\Big](E-j\omega)
+\frac{1}{B(E^2+AD)^2}\Big[W_{1}+\beta W_{1}^3\Big]\nu_{1},\nonumber\\
K_{20}&=&\frac{D}{C(E^2+AD)}\Big[(E-j\omega)J+\beta (E-j\omega)J^3\Big]+
\frac{E}{C(E^2+AD)}\Big[(E-j\omega)+\beta (E-j\omega)^3\nu_{1}^2\Big],\nonumber\\
K_{22}&=&\frac{D}{C(E^2+AD)}\Big[(E-j\omega)^2
+\beta (E-j\omega)^4\Big]-\frac{1}{BC}+\frac{E}{C(E^2+AD)}
\Big[(E-j\omega)+\beta (E-j\omega)^3\Big]-\frac{m^2}{C}\nonumber\\
&-&\frac{A}{C(E^2+AD)}\Big[\nu_{1}^2+
\beta \nu_{1}^4\Big],\nonumber\\
K_{30}&=&\frac{(AD-{A^2})}{(E^2+AD)^2}\Big[\nu_{1}+\beta \nu_{1}^3\Big](E-j\omega)+ \frac{E}{C(E^2+AD)}\Big[J^2+\beta J^4\Big]
-\frac{m^2E}{(E^2+AD)},\nonumber\\
K_{23}&=&\frac{A}{C(E^2+AD)}\Big[J+\beta J^3\Big]\nu_{1},~~~~~K_{31}=\frac{1}{B(E^2+AD)}\Big[\nu_{1}+\beta \nu_{1}^3\Big]W_{1},\nonumber\\
K_{32}&=&\frac{E}{C(E^2+AD)}\Big[J+\beta J^3\Big](E-j\omega)+
\frac{A}{C(E^2+AD)}\Big[\nu_{1}+\beta \nu_{1}^3\Big]J,\nonumber\\
K_{33}&=&\frac{(AD-{A^2})}{(E^2+AD)}\Big[(E-j\omega)^2
+\beta (E-j\omega)^4\Big]-\frac{1}{B(E^2+AD)}\Big[W_{1}^2+\beta W_{1}^4\Big]
-\frac{A}{C(E^2+AD)}\Big[J^2+\beta J^4\Big]\nonumber\\
&-&\frac{m^2 A}{(E^2+AD)},\nonumber
\end{eqnarray}
where $J=\partial_{y}S_{0}$, $W_{1}=\partial_{r}{S_{0}}$ and $\nu_{1}=\partial_{\theta}{S_{0}}$.
In order to get a significant imaginary solution, we put the determinant of matrix $\big|\textbf{K}\big|=0$, so
\begin{eqnarray}\label{a1}
ImW^{\pm}&=&\pm \int\sqrt{\frac{(E-j\omega)^{2}
+Z_{1}\Big[1+\beta\frac{Z_{2}}{Z_{1}}\Big]}{\frac{(E^2+AD)}{BD}}}dr,\nonumber\\
&=&\pm i\pi\frac{(E-j\omega)+\Big[1+\beta\Xi\Big]}{2\kappa(r_{+})},
\end{eqnarray}
here
\begin{eqnarray}
Z_{1}&=&\frac{BE(E-j\omega)\nu_{1}}{(E^2+AD)}+\frac{AB \nu_{1}^2}{(E^2+AD)}-Bm^2,\nonumber\\
Z_{2}&=&\frac{BD(E-j\omega)^4}{(E^2+AD)}+\frac{BE(E-j\omega)^3}{C(E^2+AD)}
-\frac{AB\nu_{1}^4}{(E^2+AD)}-W_{1}^4.\nonumber
\end{eqnarray}
The charged particles tunneling probability is defined as
\begin{equation}
\Gamma=\frac{\Gamma_{emission}}{\Gamma_{absorption}}=
\exp\Big[{-2\pi}\frac{(E-j\omega)}
{\kappa(r_{+})}\Big]\Big[1+\beta\Xi\Big].
\end{equation}
here
\begin{equation}
\kappa(r_{+})=\frac{2r_+\Big(2m l^2+r_+^3\Big)+6mr_+^6-8mrr_+^3\Big(2m l^2 +r_+^3\Big)}
{2\Big(2ml^2+r_+^3\Big)^2\Big(r_+^2+a^2\Big)}.
\end{equation}
The generalized $T'_{H}$ of regular rotating Hayward BH can be computed by considering the Boltzmann factor
$\Gamma_{B}=\exp\left[(E-j\omega)/T'_{H}\right]$ as follows
\begin{eqnarray}
T'_{H}&=&\frac{2r_+\Big(2m l^2+r_+^3\Big)^2+6mr_+^6-8mr_+^3\Big(2m l^2 +r_+^3\Big)}
{4\pi\Big(2ml^2+r_+^3\Big)^2\Big(r_+^2+a^2\Big)}\Big[1-\beta\Xi\Big].\label{v2}
\end{eqnarray}

The $T'_{H}$ of BH depends upon $\beta$, $m$, $a$ and $r_+$.
We also conclude that the $T'_{H}$ of the charged
vector particles is smaller than the $T_{H}$. So, quantum corrections decelerate the increase in $T_{H}$.

Equating with the $T'_{H}$, the GUP-modified particle energy radiate in regular rotating Hayward BH \cite{x17, E}
is computed as

\begin{equation}
E_{GUP}\geq E[1-\beta\Xi]
\end{equation}
the $E_{GUP}$ relates on arbitrary parameter $\Xi$ and  correction parameter $\beta$. The $E_{GUP}$ increases with the increasing values of correction parameter.

It is important to mention here that if we put ($\beta=0$), we get the original Hawking temperature $T_H$ of Eq. (\ref{bb}) for regular rotating Hayward BH. When $l=0, \beta = 0$, we obtain
the Hawking temperature of Kerr BH \cite{n1}. Moreover, for $a=0, l=0, \beta=0$, we recover the temperature of Schwarzschild BH \cite{n2}.

\section{Graphical Analysis of Corrected Temperature $T'_{H}$ for Regular Rotating Hayward BH}
This section provide the graphical analysis of $T'_H$
via horizon $r_+$ for regular rotating Hayward BH. We study the effects of quantum correction parameter $\beta$, spin parameter
$a$ and free parameter $l$ on $T'_{H}$. We analyze the behavior of $T'_{H}$ for fixed
value of mass $m=1$ and varying values of different parameters.

\begin{center}
\includegraphics[width=8cm]{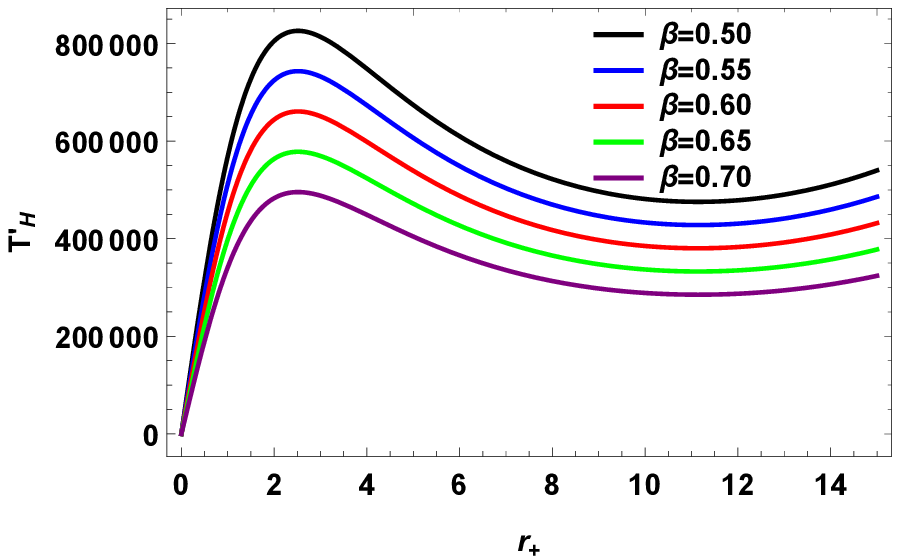}\includegraphics[width=8cm]{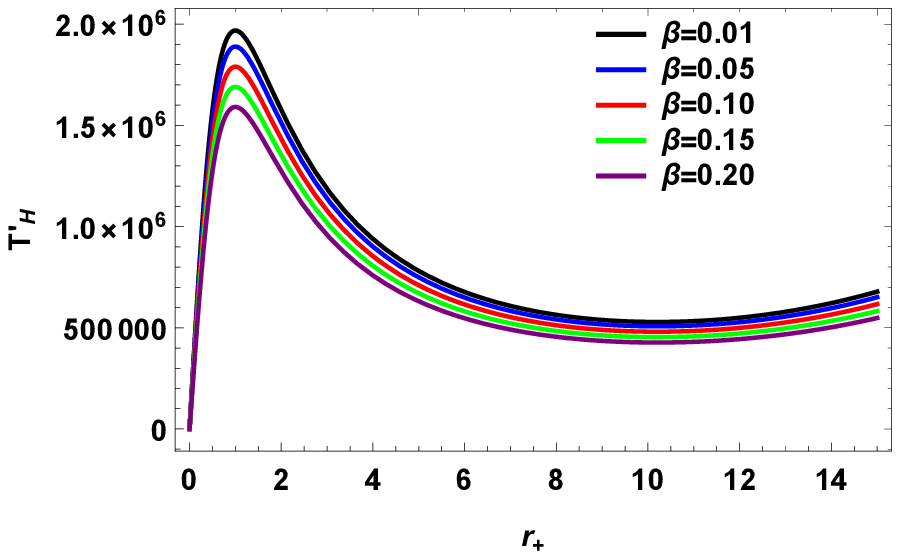}\\
{Figure 3: Hawking temperature $T'_{H}$ versus horizon $r_{+}$.}
\end{center}

\textbf{Fig. 3}: states the behavior of $T'_{H}$ for fixed values of spin parameter
$a$, free parameter $l$ and different values of correction parameter $\beta$ in the range $0\leq r_+\leq15$.

The left hand side plot shows the behavior of $T'_{H}$ for fixed values of $a=2.5,~l=60$ and varying values of gravity parameter $\beta$.
The temperature for these values of $\beta$ satisfies the GUP relation with positive temperature and shows the stable condition of BH.

The right hand plot depicts the behavior of $T'_{H}$ for fixed values of $a=1,~l=50$ for different values of correction parameter.
One can observe that the $T'_{H}$ decreases as horizon increases and remnant left at maximum temperature with non-zero horizon.
The $T'_{H}$ decreases with the increasing values of correction parameter.

\begin{center}
\includegraphics[width=8cm]{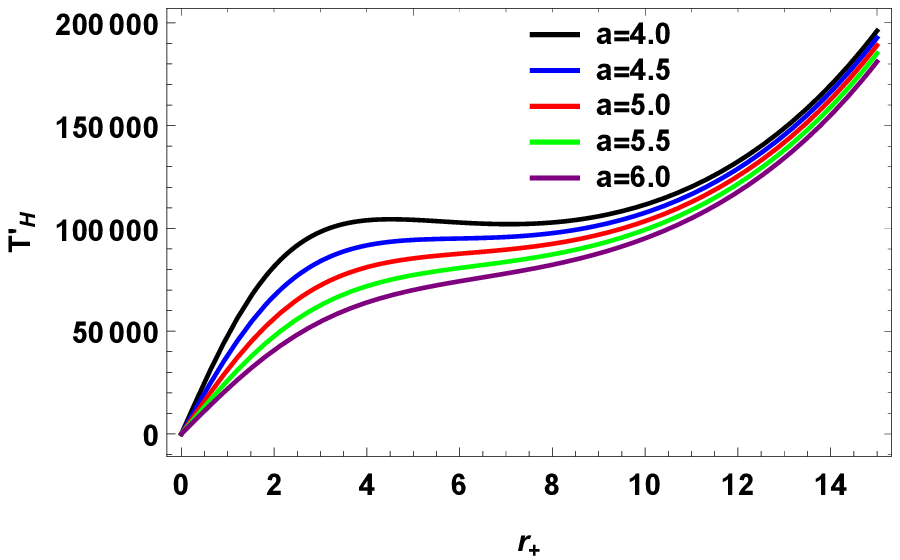}\includegraphics[width=8cm]{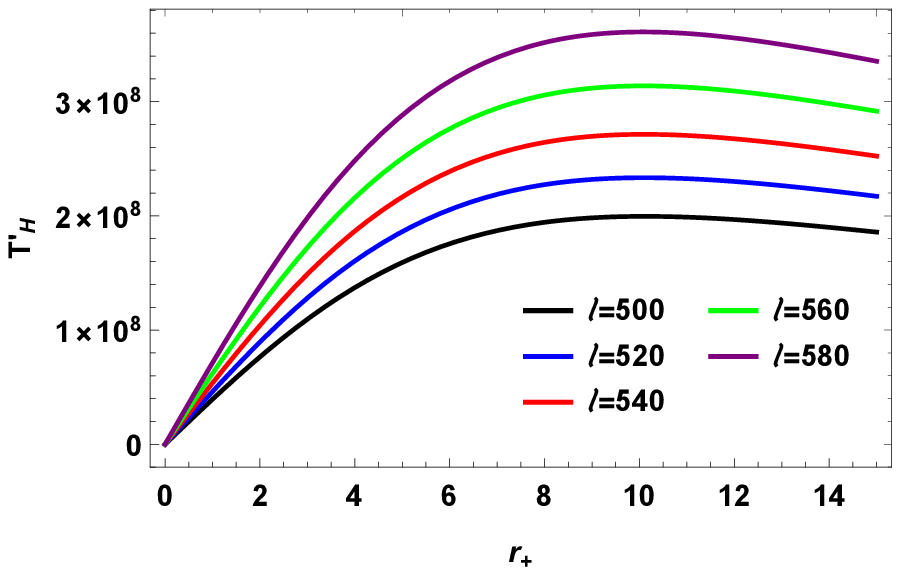}\\
{Figure 4: Hawking temperature $T'_{H}$ versus horizon $r_{+}$.}
\end{center}

\textbf{Fig. 4}: describes the graphical presentation of $T'_{H}$ for fixed values of
correction parameter and changing values of $a$ and $l$.

The left hand side plot gives the presentation of $T'_{H}$ for fixed value of $\beta=0.5, l=40$
and changing values of $a$. We can also observe that in the presence of $\beta$
influences the $T'_{H}$ also decreases with the increasing values of $a$. The presentation shows the stable condition of BH.

The right hand side plot gives the presentation of $T'_{H}$ for constant value of $\beta=0.9, a=10$
and changing values of $l$. We can see after a height the $T'_{H}$ decreases as $r_+$ increases and
temperature increases with the increasing values of $l$ as compared to absence of quantum gravity effects case.
This representation ensures the stability of BH in the range $0\leq r_+\leq15$.

It has also worth to mention here that in the case of $\beta$ effects, we get the smaller $T'_{H}$.
So, \textbf{Fig. 3 \& 4} shows that the $\beta$ effects decelerate the increase in $T'_{H}$.

\section{Logarithmic Entropy Correction for Regular Rotating Hayward BH}
In this section, we compute the entropy corrections
for regular rotating Hayward BH. Banerjee and Majhi \cite{x32}-\cite{x34} have calculated
the corrected temperature and entropy by considering the back-reaction effects with the help of null
geodesic method. We investigate the entropy corrections for regular rotating Hayward BH by utilizing the
generic formula for first order corrections to Bekenstein-Hawking formula \cite{x35}.
We calculate the logarithmic entropy corrections with the help of
corrected temperature $T'_{H}$ and standard entropy $S_{o}$ for regular rotating Hayward BH.
The entropy corrections can be calculated by the given formula
\begin{equation}
S=S_{o}-\frac{1}{2}\ln\Big|T_H^2, S_{o}\Big|+...~.\label{vv}
\end{equation}
The standard entropy for regular rotating Hayward BH can be computed by the general formula
\begin{equation}
S_{o}=\frac{A_+}{4},
\end{equation}
here
\begin{eqnarray}
A_+&=&\int_{0}^{2\pi}\int_{0}^{\pi}\sqrt{g_{\theta\theta}g_{yy}}d\theta dy,\nonumber\\
&=&\frac{2\pi\Big(r^2+a^2\Big)\Big(r^3+2l^2m\Big)}{\Big(2l^2m+r^3+mr^4\Big)}.
\end{eqnarray}
So the standard term for entropy is given as
\begin{equation}
S_{o}=\frac{\pi\Big(r^2+a^2\Big)\Big(r^3+2l^2m\Big)}{2\Big(2l^2m+r^3+mr^4\Big)}.\label{v1}
\end{equation}
After putting the values from Eq. (\ref{v2}) and (\ref{v1}) into Eq. (\ref{vv}),
we calculate the corrected entropy in the following form
\begin{equation}
S=\frac{\pi\Big(r^2+a^2\Big)\Big(r^3+2l^2m\Big)}{2\Big(2l^2m+r^3+mr^4\Big)}-\frac{1}{2}\ln\left|\frac{\left[\left(2r_+\Big(2m l^2+r_+^3\Big)^2+6mr_+^6-8mr_
+^3\Big(2m l^2 +r_+^3\Big)\right)\Big(1-\beta\Xi\Big)\right]^2}
{32\pi\Big(2ml^2+r_+^3\Big)^3\Big(r_+^2+a^2\Big)\Big(2l^2m+r^3+mr^4\Big)}\right|+...
\end{equation}
The above equation gives the corrected entropy for regular rotating Hayward BH.

\section{Conclusions}

In this paper, we have effectively applied two different phenomena
(the Newman-Janis algorithm and the semiclassical Hamilton-Jacobi method) to
analyze the $T_{H}$ for massive $4$-dimensional regular Hayward BH with spin parameter.
By considering the Newman-Janis algorithm, the Hayward BH solution
with spin parameter is computed. We have derived the temperature for regular rotating
Hayward BH with the help of surface gravity. The temperature depends upon the BH mass $m$,
spin parameter $a$, free parameter $l$ and BH radius $r_+$.
We have also analyzed the effects of spin parameter $a$ and free parameter $l$ on Hawking
temperature with the help of graphs. The temperature decreases with the increasing values of spin parameter
$a$ and increases with the increasing values of free parameter $l$.

To study the Hawking temperature of massive vector particles, we have utilized the Lagrangian
filed equation in the background of GUP within the concept of WKB approximation and semiclassical
Hamilton-Jacobi method. By taking into account the
Hamilton-Jacobi phenomenon and the complex path integration, we have calculated
the tunneling probability of the regular rotating Hayward BH, which is dominated by the well
known Boltzmann factor. Then, the temperature found from the leading tunneling
probability is established to the $T'_{H}$ of the regular rotating Hayward BH.
The $T'_{H}$ of regular rotating Hayward BH depends upon correction parameter $\beta$, BH mass $m$,
spin parameter $a$, free parameter $l$ and BH radius $r_+$.
If ($\beta=0$), we analyzed the
original $T_H$ of Eq. (\ref{bb}) for regular rotating Hayward BH. When $l=0, \beta = 0$, we found
the Hawking temperature of Kerr BH. Moreover, for $a=0, l=0, \beta=0$, we observed the temperature of Schwarzschild BH.
Moreover, we have studied the graphical
presentation of  $T'_H$ via horizon $r_+$ for regular rotating Hayward BH.
We have investigated the effects of quantum
gravity $\beta$, spin parameter $a$ as well as free parameter $l$ on corrected temperature. We have concluded that, we observed an increasing behavior of temperature for the increasing values of $l$ and decreasing behavior for increasing values of $a$ and $\beta$.
The temperature at maximum height with non-zero horizon gives BH remnant. After maximum height
the temperature eventually goes down and get an asymptotically flat condition till $r_+\rightarrow\infty$,
which ensures the stable case of BH. Comparing with the $T'_{H}$, the GUP-modified particle energy radiate in regular rotating Hayward BH also computed.
The particle energy $E_{GUP}$ radiate in regular rotating Hayward BH depends on quantum gravity.
 
Finally, we study the entropy corrections for regular rotating Hayward BH by observing the generic formula for first order corrections to Bekenstein-Hawking formula \cite{x35} and analyze the logarithmic entropy corrections with the help of $T'_{H}$ and standard entropy $S_{o}$ for regular rotating Hayward BH.

\section*{Acknowledgments}

We are very much grateful to the honorable referee and to the editor for the
illuminating suggestions that have significantly improved our work in terms
of research quality, and presentation.

\end{document}